\documentclass[12pt]{article}

\usepackage{latexsym}
\usepackage{graphicx}
\usepackage{rotating}

\def\Rmath{\mbox{\hbox{\it I\hskip -2.2pt R}}}

\def\emath{\mbox{e}}

\title{Multiple Resonances in Fluid-Loaded Vibrating Structures}
\author {{\sc P.-O.~Mattei} \\
{\sl Laboratoire de M\'ecanique et d'Acoustique} \\
{\sl 31 chemin Joseph Aiguier} \\
{\sl 13402 Marseille Cedex 20 (France) }  }
\date{}

\begin{document}

\maketitle

%==================================================================
 
\noindent{\bf Summary}\\
This study deals with spectral analysis of fluid-loaded vibrating structure. It was recently observed~\cite{POM-2007} in a numerical study on a high order perturbation method under heavy fluid loading that a loaded vibrating plate results, not only in the classical  frequency shift of the in vacuo single resonance (in both the real part because of the fluid added mass and the imaginary part because of energy lost by radiation into the fluid), but also in an increase in the number of the resonance frequencies : as a result of the loading, a single in vacuo resonance frequency of the structure is transformed into a multiple resonance frequency. Here we show that this phenomenon is said to be a refinement of the Sanchez's classical result~\cite{Sanchez-Sanchez-1989}(paragraph 9.3) where it was established, using asymptotic analysis, that in the case of a light loading conditions ``the scattering frequencies of a fluid loaded elastic structure (ie the resonance frequencies) are nearly the real eigenfrequencies of the elastic body alone and the complex scattering frequencies of the fluid with a rigid solid''.
          
A theoretical explanation of the observed phenomenon of multiple resonance is given using classical results of the distribution of zeros of entire functions. It is established that every single in vacuo resonance frequency of a rectangular plate is transformed into an infinite number of resonances when the fluid-loading is accounted for. 

%==================================================================
 
\section{Introduction}

It is  well known that the need to accurately describe fluid structure interactions increases the computational cost of the various numerical methods used. Most numerical methods can be used to deal with the resolution of linear systems of simultaneous equations. When the loading conditions are taken into account, the size of the matrices involved increases considerably, and they become full and frequency dependent. Any method of reducing these drawbacks is therefore most welcome; and one of the best methods available is that based on the perturbation approach. Asymptotic analysis can be conduced when the loading is light, as in the case of a metal structure surrounded by air. This approach consists in introducing a small parameter $\epsilon$, which is the ratio between the surface mass density and the fluid density. Using perturbation expansions~\cite{Nayfeh}, one can then construct an approximate solution, based on the {\it in vacuo} eigenmodes, which is not only easier to calculate than the solution of the exact problem but also leads to a better understanding of the various phenomena involved~\cite{PJTF-DH-POM-CM}. 

The light fluid loading perturbation method (involving high density structures in contact with a small density fluid) was recently extended~\cite{POM-2007} to cases where the perturbation parameter becomes large (which corresponds to a light and/or thin structure in contact with a high density fluid). During this study multiple resonance processes were observed where a single resonance mode can have several resonance frequencies. The spectrum of the operator can no longer be described in terms of the discrete spectrum based on a countable series of resonance mode/resonance frequency pairs. This typically nonlinear phenomenon is closely linked to the concept of non-linear frequency modes (\cite{Dazel-Lamarque-Sgard}), where the non linearity depends on the time parameters (such as those involved in porous materials) rather than on the more usual geometrical parameters. This frequency non-linearity is related to problems which are written in the form ${\cal O} (\omega) U = S$, where $u$ is the unknown factor and $S$ is the source term.  ${\cal O} (\omega)$ is an operator which depends non linearly on the angular frequency $\omega$, as in the case of porous environments or in the context of vibroacoustics, where the coupling depends non-linearly on the frequency via the Helmholtz equation Green's kernel.

Here it is established that this behavior is a general property of at least fluid-loaded simply supported rectangular plates. The main results obtained in our previous study~\cite{POM-2007} are recalled in paragraph~\ref{par:statement}; in particular, the method for calculating the spectrum of a fluid loaded vibrating structure based on a high order perturbation expansion is recalled and the resonance frequency equation to be solved is given. A theoretical description of multiple resonance in the case of a simply supported plate loaded with a non vanishing density fluid is given in paragraph~\ref{par:multiple}. It is then established, using classical results on the connection between the order of an entire function and the distribution of its zeros, that any resonance mode in this structure will have an infinite number of resonance frequencies. Numerical examples of a steel plate in contact with air or water are given  paragraph~\ref{par:num-examples}. The conclusions and the possible extensions of this work are presented in paragraph~\ref{par:conclusions}.

%==================================================================
 
\section{Statement of the problem} \label{par:statement}

Consider a finite elastic structure occupying a domain $\Sigma$, the behavior of which is described by a differential operator ${\cal A}$ (in the case treated in this communication it is a simple bi-Laplacian operator). This structure is loaded with a perfect fluid at rest extending to infinity. The equation of the fluid-loaded structure is given by the classical integrodifferential equation
\begin{equation}
{\cal A} U(M,\omega) - \rho_s h \omega^2 \left( U(M,\omega) - \epsilon \int_{\Sigma}  U(M',\omega) G(M;M',\omega) dM' \right) = F(M) \label{eq:strong-formulation}
\end{equation}
where $\epsilon = \rho_f/\rho_s h$ is a small parameter in the case of a metal structure (volume mass density $\rho_s$ and thickness $h$) in contact with air (volume mass density $\rho_f$). $G(M;M',\omega)$ is the Green's function of the Neumann problem in the Helmholtz equation outside the surface $\Sigma$. $F(M)$ is the source. To facilitate reading, the weak formulation of the problem is introduced.  One defines the radiation impedance $\beta (U,V,\omega) = \int_{\Sigma} \int_{\Sigma}  U(M') G(M;M',\omega) V^*(M) d M d M'$, where $U(M)$ and $V(M)$ are functions defined on the surface $\Sigma$, $a(U,V^*)$ stands for the potential elastic energy of the structure, $\langle U,V^* \rangle$ is the usual inner product and  $\rho_s h \omega^2 \langle U,V^* \rangle$ is the kinetic energy. Then the weak formulation reads: find $U(M)$ such that for every $V(M)$ one has 
\begin{equation}
a(U,V^*) - \tilde{\Lambda}_m(\omega) \left( \langle U,V^* \rangle - \epsilon  \beta(U,V^*,\omega) \right) =  \langle F,V^* \rangle. \label{eq:weak-formulation}
\end{equation}

\subsection{Resonance modes and eigenmodes for a fluid-loaded structure} 

The eigenmodes $\tilde{U}_m(M,\omega)$ and eigenvalues $\tilde{\Lambda}_m(\omega)$ (or equivalently the eigenpulsations $\tilde{\omega}_m(\omega)$ given by $\tilde{\Lambda}_m (\omega) = \rho_s h \tilde{\omega}_m^2(\omega)$) are the non zero solutions of the homogeneous weak formulation~(\ref{eq:weak-formulation}), that is find $\tilde{U}_m(M,\omega)$ and $\tilde{\omega}_m(\omega)$ such that for every $V(M)$ 
\begin{equation}
a(\tilde{U}_m,V^*) - \rho_s h \tilde{\omega}_m^2(\omega) \left( \langle \tilde{U}_m,V^* \rangle - \epsilon  \beta (\tilde{U}_m,V^*,\omega) \right) =  0, \label{eq:eigen-modes}
\end{equation}
$\beta(U,V,\omega)$, that characterizes the energy loss by radiation, depends on the frequency. Then the eigenmodes, well suited to deal with harmonic regime, are complex and depend on the frequency.

One introduces the resonance modes, that allow to compute the transient regime in a natural way;  they are complex and do not depend on the frequency~\cite{PJTF-DH-POM-CM}. The resonance modes $\hat{U}_m(M)$ and resonances pulsations $\hat{\omega}_m$, given by $\hat{\Lambda}_m =\rho_s h \hat{\omega}_m^2$, are the non zero solutions of the following problem: find $\hat{U}_m(M)$ and $\hat{\omega}_m$ such that for every $V(M)$  
\begin{equation}
a(\hat{U}_m,V^*) - \rho_s h \hat{\omega}_m^2 \left( \langle \hat{U}_m,V^* \rangle - \epsilon  \beta (\hat{U}_m,V^*,\hat{\omega}_m)  \right) =  0. \label{eq:reson-modes}
\end{equation}

These two kind of modes, that are identical when the loading is not took into account, have a strong link as each resonance pulsation is the solution of the resonance equation
\begin{equation}
\tilde{\omega}_m^2 (\hat{\omega}_m) =  \hat{\omega}_m^2, \label{eq:reson-puls-equation}
\end{equation}
and each resonance mode is an eigenmode calculated at the corresponding resonance pulsation. 
\begin{equation}
\hat{U}_m (M) =  \tilde{U}_m(M,\hat{\omega}_m). \label{eq:reson-mode-equation}
\end{equation}

It is worth noting that in the eigenmodes are much more easier to compute than the resonance modes.

\subsection{High order perturbation method} 

When the loading parameter $\epsilon$ is small, the eigenmodes and eigenvalues can be calculated using a perturbation method. The following perturbation expansion is introduced into the weak formulation~(\ref{eq:eigen-modes})
\begin{eqnarray*}
\tilde{U}_m (M) & = & \tilde{U}_m^{(0)}(M) + \epsilon \tilde{U}_m^{(1)} (M) + \cdots + \epsilon^s  \tilde{U}_m^{(s)}(M) + \cdots \\
\tilde{\Lambda}_m & = & \tilde{\Lambda}_m^{(0)} + \epsilon \tilde{\Lambda}_m^{(1)} + \cdots + \epsilon^s \tilde{\Lambda}_m^{(s)} +  \cdots
\end{eqnarray*}
where $\tilde{U}_m^{(0)}(M)$ and $\tilde{\Lambda}_m^{(0)}$ are the eigenmodes and eigenvalues of the elastic structure {\it in vacuo}. To facilitate the reading, one poses: $\beta^{mn}(\omega) = \beta\left( \tilde{U}_m^{(0)}, \tilde{U}_n^{(0)*}, \omega \right)$. It can then be easily established that one has
\begin{eqnarray*}
\mbox{to the order 1 in $\epsilon$:}  \tilde{\Lambda}_m^{(1)} (\omega) & = & \tilde{\Lambda}_m^{(0)} \beta^{mm}(\omega), \\
\mbox{to the order 2 in $\epsilon$:} \tilde{\Lambda}_m^{(2)} (\omega) & = & \tilde{\Lambda}_m^{(0)} \left( (\beta^{mm}(\omega))^2 + \sum_{p \neq m} \frac{ \tilde{\Lambda}_m^{(0)} }{ \tilde{\Lambda}_m^{(0)} - \tilde{\Lambda}_p^{(0)}} \beta^{mp}(\omega) \right),
\end{eqnarray*}
and so on to the higher orders. Next, as previously shown~\cite{POM-2007}, by grouping the terms and by a re-summation technique, one shows that the high order perturbation expansion is given by 
\begin{equation}
\tilde{\Lambda}_m (\omega) \approx \frac{\tilde{\Lambda}_m^{(0)}}{1 - \epsilon \beta^{mm}(\omega)},\label{eq:high-order-expansion}
\end{equation}
which remains valid even for higher values of $\epsilon$. It is worth noting that, compared to the classical perturbation expansion of the eigenvalues given by $\tilde{\Lambda}_m (\omega) \approx \tilde{\Lambda}_m^{(0)} \left(1 +\epsilon \beta^{mm}(\omega)\right)$, the high-order approximation~(\ref{eq:high-order-expansion}) is particularly interesting as it gives very precise results for the same numerical cost.

The calculation of the eigenmodes and eigenvalues does not requires any particular effort and, for a given $\omega$, each eigenpulsation is the solution of the following eigenpulsation equation 
\begin{equation}
\tilde{\omega}_m^2 (\omega) \approx \frac{\tilde{\omega}_m^{(0)2}}{ \left(1 - \epsilon \beta^{mm} (\omega) \right)}. \label{eq:approx-eigen-puls-equation}
\end{equation}

Equation~(\ref{eq:approx-eigen-puls-equation}) shows that the calculation of the various eigenpulsations do not requires any computational cost but the evaluation of multiple integrals $\beta^{mm} (\omega)$. 

\subsection{The resonance equation} 

By introducing in equation~(\ref{eq:reson-puls-equation}) the high order expansion of the eigenpulsation given by equation~(\ref{eq:approx-eigen-puls-equation}), the resonance equation~(\ref{eq:reson-puls-equation}) is approximated as
\begin{equation}
\hat{\omega}_m^2 \left(1 - \epsilon \beta^{mm} (\hat{\omega}_m) \right) \approx \tilde{\omega}_m^{(0)2}. \label{eq:approx-reson-puls-equation}
\end{equation}

This shows that the calculation of the resonance pulsations is completely different  than that of the eigenpulsations since $\beta^{mm} (\omega)$ has a rather complicated dependence on the frequency and solving equation~(\ref{eq:approx-reson-puls-equation}) requires the search for the zeros of non-convex complex function which is a challenging task. A numerical example given in the paper~\cite{POM-2007} has shown a curious phenomenon of multiple resonance. It has been made a comparison of the evolution of the resonance frequency in the mode $(1-3)$ of clamped steel plate loaded by water when thickness vary using an exact solution and the approximated one (by the high order perturbation expansion). That particular example has showed that the resonance frequency is not unique and at least two resonance frequencies could exist simultaneously for a given geometrical configuration. 

To have a better understanding of this non-usual phenomenon the roots of the resonance equation~(\ref{eq:approx-reson-puls-equation}) are now studied in a more general manner using classical results  in the framework of the distribution of zero of entire functions\cite{Levin-1978, Levin-1996}. In the following the resonance equation~(\ref{eq:approx-reson-puls-equation}) is written as $f^{mm}(z)$:
\begin{equation}
f^{mm}(z) = z^2 \left(1 - \epsilon \beta^{mm} (z) \right) - z_{0m}^2, \label{eq:reson-function}
\end{equation}
where $z = \hat{\omega}_m$ and $z_{0m}= \tilde{\omega}_m^{(0)}$. The zeros of this function are the resonance pulsations that are looked for.

%==================================================================
 
\section{Multiple resonances} \label{par:multiple}

\subsection{Methodology}

Now all the problem focuses on the existence and the counting of the roots of $f^{mm}(z) = z^2 \left(1 - \epsilon \beta^{mm} (z) \right) - z_{0m}^2$. Let us presents the methodology used in the rest of paper. More or less, the theory of the zeros of entire functions (analytic functions in the whole complex plane) is based on the Great Picard's Theorem~\cite{Saks-Zygmund-1965} which states that if $f(z)$ has an essential singularity at a point $z_w$ then on any open set containing $z_w$, $f(z)$ takes on all possible values, with at most one exception, infinitely often. The exception is needed as showed by the function $\exp(1/z)$ that has an essential singularity at $z=0$ but never attains $0$ as a value. The second point is that every entire function is either a polynomial or has an essential singularity at infinity. Now if one can prove that the resonance function $f^{mm}(z)$ is an entire function, that is not a polynomial and that the value $0$ is not the exception needed by the Picard's theorem, then the resonance function $f^{mm}(z)$ has an infinite number of zeros. Then for every mode, an infinite number of resonance exists. 

Now to characterize complex function $f(z)$ as an entire function, the simplest way is to study its development as a power series of the form $f(z) = \sum_{n=0}^{n=\infty} c_n z^n$. When $\lim_{n\rightarrow \infty} \sqrt[n]{c_n} = 0$, $f(z)$ is an entire function. To show that it is not a polynomial, one can study its order. Let us recall that if $f(z)$ is an entire function, the function $M(r) = \max_{|z| = r} |f(z)|$ will increase indefinitely together with $r$. The rate of growth of $M(r)$ can be characterized by comparing it with function $\exp(r)$. By definition, the order (of growth) $\rho$ is the smallest number such that $M(r) \le \exp r^{\rho+\eta}$ for every $\eta >0$ and for $r > r_0(\eta)$. From this, it can be easily seen that the order is given by the formula $\rho = {\lim \sup}_{r \rightarrow \infty} \ln \ln M(r) / \ln r$ ; as examples, a polynomial in $z$ is of order 0, and the exponential function is of order 1. The order $\rho$ of an entire function depends on the asymptotic behavior of the coefficients of its series expansion. One has $1/\rho = \lim_{n\rightarrow \infty} \inf (\ln (1/|c_n|) / (n \ln n))$ (\cite{Saks-Zygmund-1965}); or in other words, if $c_n \propto 1/(n!)^{\alpha}$ then $\rho = 1/\alpha$. 

One can simplify again the problem under consideration. The resonance function can be written as $f^{mm}(z) = z^2 - z_{0m}^2 - \epsilon z^2  \beta^{mm} (z)$, as a sum of a polynomial $(z^2 - z_{0m}^2)$ of order 0 and a function which is the product of a polynomial $(\epsilon z^2)$ also of order 0 and the radiation impedance $\beta^{mm} (z)$ which order is to be determined. The distribution of zeros of $f^{mm}(z)$ can be deduced from that of $\beta^{mm} (z)$ as long as $\epsilon$ has a non zero value. If $\beta^{mm} (z)$ is a polynomial, the resonance function is also a polynomial with a finite number of roots. If $\beta^{mm} (z)$ is an entire function  (in the case considered below it is of order 1) with essential singularity at infinity, then $f^{mm}(z)$ is also an entire function with essential singularity at infinity and by the Picard's theorem $f^{mm}(z)$ can takes on all possible values at infinity, with at most one exception, infinitely often. 

\hspace{1ex}

The main results of this paper is, as shown below on the particular example of a simply supported plate, that the radiation impedance and the resonance function are entire functions of exponential type (that is of order one and normal type~\cite{Levin-1978}) that both have an infinite number of roots.

\subsection{The resonance function of a simply supported baffled plate as an entire function of the frequency}

Now let us consider a thin simply supported rectangular plate (dimensions $a \times b$ with thickness $h$) made of isotropic material with Young's modulus $E$, Poisson's coefficient $\nu$ and density $\rho_s$.  In a vacuum, the eigenmodes are given by $\tilde{U}_{mn}^{(0)}(x,y) = 2/\sqrt{ab} \sin m \pi x /a \sin n \pi y /b$ and the eigenpulsations by $\omega_{mn}^{(0)} = \pi^2 \sqrt{D/(\rho_s h)} \left[(m/a)^2 + (n/b)^2\right]$ with $D=E h^3 /12(1-\nu^2)$. When the plate is baffled, the radiation impedance $\beta ( \tilde{U}_{mn}^{(0)}, \tilde{U}_{mn}^{(0)*}, \omega )$ is given as~\cite{Nelisse-Beslin-Nicolas}
\begin{equation}
\beta ( \tilde{U}_{mn}^{(0)}, \tilde{U}_{mn}^{(0)*}, \omega ) = - \frac{a b }{\pi} \int_0^1 \int_0^1 \int_0^1 \int_0^1 \sin m \pi x \sin n \pi y  \sin m \pi x' \sin n \pi y' \frac{e^{\imath \frac{\omega}{c} d}}{ d} d x d y d x' d y',
\end{equation}
where $d=\sqrt{a^2(x-x')^2+b^2(y-y')^2}$ and $c$ is the speed of waves in the fluid. In the following, the pulsation dependence in $\omega$ is changed into the equivalent wavenumber dependence in $k = \omega/c$. Since the in vacuo modes involve a double index, the notation for the radiation impedance is simplified as $\beta ( \tilde{U}_{mn}^{(0)}, \tilde{U}_{mn}^{(0)*}, \omega ) = \beta^{mn}(k)$. The resonance function is denoted by $f^{mn} (k) = k^2 (1- \epsilon \beta^{mn}(k)) - k_{0mn}^2$, with $k_{0mn}= \omega_{mn}^{(0)}/c$.

It is worth noting that $\beta^{mn} (k)$ and $f^{mn} (k)$ decrease in the upper complex plane (with $\Im (k) > 0$, the term $\exp(\imath k d)$ has an exponential decrease when $|k|\rightarrow \infty$) and conversely grow in the lower complex plane, then all the roots of these two functions are located in the lower complex half plane. As the time dependency had been chosen as $\exp( -\imath \omega t)$, this ensures that the movement of the plate remains finite over time~\cite{PJTF-DH-POM-CM}. 

By making a change of variable~\cite{Mangiarotty}, it can be easily shown that $\beta^{mn}(k)$ is given by 
\begin{equation}
\beta^{mn} (k) = - \frac{4 a b }{\pi} \int_0^1 \int_0^1 G_m (X) G_n (Y) \frac{e^{\imath k D}}{D} d X d Y, \label{eq:rad-impedance}
\end{equation}
where $D = \sqrt{a^2 X^2+b^2Y^2}$. The functions $G_m (x)$ and $G_n (y)$ are given by single analytic integrals $G_m (x) = \int_0^1 \sin m \pi (x+x') \sin m \pi x' d x'$ with $G_m (x) =  \left( \sin m \pi x + m \pi(1-x) \cos m \pi x \right)/(2 m \pi)$. By making a change of variables from rectangular to polar~\cite{DH-PJTF-2004}, with $0<\theta_{\alpha} = \arctan(b/a)<\pi/2$, the radiation impedance~(\ref{eq:rad-impedance}) is written as
\begin{equation}
\beta^{mn} (k) = -\frac{4 }{\pi} \int_0^{\theta_{\alpha}} J_{1mn}(k,\theta) d \theta  - \frac{4 }{\pi} \int_{\theta_{\alpha}}^{\pi/2} J_{2mn}(k,\theta) d \theta, \label{eq:rad-impedance-polar}
\end{equation}
with 
\begin{eqnarray}
J_{1mn}(k,\theta) & = & \int_0^{\frac{a}{\cos \theta}} H_{mn} (R,\theta) e^{\imath k R} d R, \label{eq:J1mn} \\
J_{2mn}(k,\theta) & = & \int_0^{\frac{b}{\sin \theta}} H_{mn} (R,\theta) e^{\imath k R} d R, \label{eq:J2mn} \\
H_{mn} (R,\theta)  & = & G_m \left( \frac{R \cos \theta}{a} \right) G_n \left( \frac{R \sin \theta}{b} \right). \label{eq:Hmn}
\end{eqnarray}

By expanding the exponential as a power series, the integrals (\ref{eq:J1mn}) and (\ref{eq:J2mn}) are given as
\begin{eqnarray*}
J_{1mn}(k,\theta) & = &  \sum_{\ell =0}^{\ell=\infty} \frac{(\imath k)^{\ell} }{\ell !} \int_0^{\frac{a}{\cos \theta}} H_{mn} (R,\theta) R^{\ell} d R,\\
J_{2mn}(k,\theta) & = & \sum_{\ell =0}^{\ell=\infty} \frac{(\imath k)^{\ell} }{\ell !}  \int_0^{\frac{b}{\sin \theta}} H_{mn} (R,\theta) R^{\ell} d R,
\end{eqnarray*}
as $H_{mn} (R,\theta) R^{\ell}$ is a continuous integrable function of $R \in ]0,a/\cos \theta[$ and $R	\in ]0,b/\sin \theta[$, the mean value theorem ensures that there exist $R_1	\in ]0,a/\cos \theta[$  and $R_2	\in ]0,b/\sin \theta[$ such that 
\begin{eqnarray*}
\int_0^{\frac{a}{\cos \theta}} H_{mn} (R,\theta) R^{\ell} d R = H_{mn} (R_1,\theta) R_1^{\ell},\\
\int_0^{\frac{b}{\sin \theta}} H_{mn} (R,\theta) R^{\ell} d R = H_{mn} (R_2,\theta) R_2^{\ell}.
\end{eqnarray*}
Similarly as $H_{mn} (R,\theta)$ is a continuous integrable function of $\theta \in ]0,\pi/2[$, there exist $\theta_1 \in ]0,\theta_{\alpha}[$  and $\theta_2 \in ]\theta_{\alpha}, \pi/2[$ such that
\begin{eqnarray*}
\int_0^{\theta_{\alpha}} J_{1mn}(k,\theta) d \theta  = \sum_{\ell =0}^{\ell=\infty} \frac{(\imath k)^{\ell} }{\ell !} H_{mn} (R_1,\theta_1) R_1^{\ell},\\
\int_{\theta_{\alpha}}^{\pi/2} J_{2mn}(k,\theta) d \theta  = \sum_{\ell =0}^{\ell=\infty} \frac{(\imath k)^{\ell} }{\ell !}   H_{mn} (R_2,\theta_2) R_2^{\ell}.
\end{eqnarray*}
One then obtains:
\begin{equation}
\beta^{mn} (k) = -\frac{4 }{\pi} \left( H_{mn} (R_1,\theta_1) \sum_{\ell =0}^{\ell=\infty} \frac{(\imath k R_1)^{\ell} }{\ell !} +H_{mn} (R_2,\theta_2) \sum_{\ell =0}^{\ell=\infty} \frac{(\imath k R_2)^{\ell} }{\ell !}  \right),
\end{equation}
$\beta^{mn} (k)$ is then an entire function of the variable $k$.

\subsubsection{Order of the radiation impedance}

To calculate the order of the impedance radiation, one studies the comportment of the function $M_{\beta^{mn}}(r) = \max_{|k| = r} |\beta^{mn} (k)|$ for $r\rightarrow \infty$. Let us begins with successive integrations by parts of the integrals given by equations (\ref{eq:J1mn}) and (\ref{eq:J2mn}). Introducing the results given in the appendix leads to 
\begin{eqnarray}
J_{1mn}(k,\theta) & = & -\frac{1}{4 \imath k } + \frac{1}{\left(\imath k\right)^3} h_{2mn} (\theta) - \frac{1}{\left(\imath k\right)^4} \left( h_{3mn}^1 (\theta) \emath^{\imath k \frac{a}{\cos \theta}} - h_{3mn}^0 (\theta)\right) \nonumber \\
& &  + \frac{1}{\left(\imath k\right)^5} \int_0^{\frac{a}{\cos \theta}} \frac{\partial^4 H_{mn} (R,\theta)}{\partial R^4} e^{\imath k R} d R, \label{eq:J1mn-by-parts} \\
J_{2mn}(k,\theta) & = &  -\frac{1}{4 \imath k } + \frac{1}{\left(\imath k\right)^3} h_{2mn} (\theta) - \frac{1}{\left(\imath k\right)^4} \left( h_{3mn}^2 (\theta) \emath^{\imath k \frac{a}{\cos \theta}} - h_{3mn}^0 (\theta)\right) \nonumber \\
& &  + \frac{1}{\left(\imath k\right)^5} \int_0^{\frac{b}{\sin \theta}} \frac{\partial^4 H_{mn} (R,\theta)}{\partial R^4} e^{\imath k R} d R, \label{eq:J2mn-by-parts} 
\end{eqnarray}
As one studies the comportment of $\max_{|k| \rightarrow \infty} |\beta^{mn} (k)|$, in the equations~(\ref{eq:J1mn-by-parts}) and~(\ref{eq:J2mn-by-parts}), the first two power terms in $k$ are negligible compared to the exponential ones when $|k| \rightarrow \infty$; similarly, only the first exponential term has a contribution when $|k| \rightarrow \infty$. Then
\begin{equation}
M_{\beta^{mn}}(r)  =  \max_{|k| = r} \frac{1}{|k|^4} \left| \int_0^{\theta_{\alpha}} h_{3mn}^1 (\theta) \emath^{\imath k \frac{a}{\cos \theta}} d \theta + \int_{\theta_{\alpha}}^{\frac{\pi}{2}}  h_{3mn}^2 (\theta) \emath^{\imath k \frac{b}{\sin \theta}} d \theta \right|. \label{eq:M(Bmn)(r)}
\end{equation}
Now by expanding the exponentials in equation~(\ref{eq:M(Bmn)(r)}) as power series, one obtains
\begin{equation}
M_{\beta^{mn}}(r)  =  \max_{|k| = r} \frac{1}{|k|^4} \left| \sum_{\ell =0}^{\ell=\infty} \frac{(\imath k)^{\ell} }{\ell !} \left( \int_0^{\theta_{\alpha}} h_{3mn}^1 (\theta) \left(\frac{a}{\cos \theta}\right)^{\ell} d \theta + \int_{\theta_{\alpha}}^{\frac{\pi}{2}}  h_{3mn}^2 (\theta) 
\left(\frac{b}{\sin \theta}\right)^{\ell} d \theta \right) \right|. \label{eq:M(Bmn)(r)-series}
\end{equation}
As $h_{3mn}^1 (\theta)$ and $a/\cos \theta$ are continuous integrable functions for $\theta \in [0, \theta_{\alpha}]$ and $h_{3mn}^2 (\theta)$ and $b/\sin \theta$ are continuous integrable functions for $\theta \in [\theta_{\alpha},\pi/2]$. One applies the mean value theorem for the two integrals that appear in equation~(\ref{eq:M(Bmn)(r)-series}): there exist two values $\theta_1$ and $\theta_2$ such that 
\begin{eqnarray}
\int_0^{\theta_{\alpha}} h_{3mn}^1 (\theta) \left(\frac{a}{\cos \theta}\right)^{\ell} d \theta & = & h_{3mn}^1 (\theta_1) \left(\frac{a}{\cos \theta_1}\right)^{\ell}, 0<\theta_1<\theta_{\alpha}, \label{eq:mean-value-1}\\
\int_{\theta_{\alpha}}^{\frac{\pi}{2}}  h_{3mn}^2 (\theta) \left(\frac{b}{\sin \theta}\right)^{\ell} d \theta  & = & h_{3mn}^2 (\theta_2) \left(\frac{b}{\sin \theta_2}\right)^{\ell}, \theta_{\alpha}<\theta_2<\pi/2.  \label{eq:mean-value-2}
\end{eqnarray}

Let us denote $c_1 = a/\cos\theta_1$ and  $c_2 = b/\sin \theta_2$. It is easy to see $a \le c_1 \le \sqrt{a^2+b^2}$ and  $b \le c_2 \le \sqrt{a^2+b^2}$. $M_{\beta^{mn}}(r)$ is then written as 
\begin{eqnarray}
M_{\beta^{mn}}(r) & = & \max_{|k|= r} \frac{1}{|k|^4} \left| h_{3mn}^1 (\theta_1) \sum_{\ell =0}^{\ell=\infty} \frac{(\imath k c_1)^{\ell} }{\ell !} + h_{3mn}^2 (\theta_2) \sum_{\ell =0}^{\ell=\infty} \frac{(\imath k c_2)^{\ell} }{\ell !} \right| \nonumber \\
 & = & \max_{|k| = r} \frac{1}{|k|^4} \left| h_{3mn}^1 (\theta_1) \emath^{\imath k c_1} + h_{3mn}^2 (\theta_2) \emath^{\imath k c_2} \right|. \label{eq:M(Bmn)(r)-order1}
\end{eqnarray}
The previous result shows that the function $M_{\beta^{mn}}(r)$, characterizing the growth of the radiation impedance, is the sum of two exponential functions divided by a polynomial of the fourth degree. Its order is easy to obtain from classical results on the sum
and product of entire functions  (see for example \cite{Saks-Zygmund-1965}, chapter VII paragraph 6). There is two possibilities for characterizing the growth of $M_{\beta^{mn}}(r)$ for $r$ sufficiently large:  the first is that if $\sigma$ denotes a finite positive number and $C$ a positive constant one has $M_{\beta^{mn}}(r) \approx C \exp( \sigma r)$ which implies that the radiation impedance is of order 1, the second is that $M_{\beta^{mn}}(r) = 0$ which implies that the radiation impedance is of order 0. From the expressions of the functions $h_{3mn}^1 (\theta)$ and $h_{3mn}^2 (\theta)$  given in the appendix, it is obvious that for $a\neq b$ and for $m\neq n$ one cannot have $h_{3mn}^1 (\theta_1) \emath^{\imath k c_1} =- h_{3mn}^2 (\theta_2) \emath^{\imath k c_2}$. If, for example $c_1 > c_2$, then
\begin{equation}
M_{\beta^{mn}}(r) =  \max_{|k| = r} \frac{1}{|k|^4} \left| h_{3mn}^1 (\theta_1) \emath^{\imath k c_1}\right|. \label{eq:M(Bmn)k-order1}
\end{equation}
If $a = b$ and $m=n$ care must be taken; in this case $\theta_{\alpha} = \pi/4$ and a little algebra shows that 
\begin{equation}
\int_0^{\frac{\pi}{4}} h_{3mm}^1 (\theta) \left(\frac{a}{\cos \theta}\right)^{\ell} d \theta = 
\int_{\frac{\pi}{4}}^{\frac{\pi}{2}}  h_{3mm}^2 (\theta) \left(\frac{a}{\sin \theta}\right)^{\ell} d \theta. \label{eq:h3mm1=h3mm2}
\end{equation}
Then the maximum of $\beta^{mm} (k)$ is  
\begin{equation}
M_{\beta^{mm}}(r) =  \max_{|k| = r} \frac{2}{|k|^4} \left| h_{3mm}^1 (\theta_1) \emath^{\imath k c_1}\right|. \label{eq:M(Bmm)k-order1}
\end{equation}

Then $\beta^{mn}(k)$ is an entire function of order 1. The resonance function $f^{mn}(k) = k^2 ( 1- \epsilon \beta^{mn} (k)) - k_{0mn}^2$ is an  also an entire function of order 1 for all $\epsilon\neq 0$.

\subsubsection{Distribution of zeros of the radiation impedance and of the resonance function} \label{par:Cartwright}

In addition to the order, the relations~(\ref{eq:M(Bmn)(r)-order1}), (\ref{eq:M(Bmn)k-order1}) and (\ref{eq:M(Bmm)k-order1}) give a more refined information on the distribution of zeros of the radiation impedance. An entire function that grows as $\exp(\sigma r)$, with $0< \sigma< \infty$, is an entire function of order one and normal type $\sigma$. Such functions are known as entire function of exponential type (EFET)~\cite{Levin-1996}. The resonance function and the radiation impedance are EFET. They both possess an additional property on the distribution of their zeros. They are functions that belong to the Cartwright class ${\cal C}$~\cite{Levin-1996}. A EFET $f(z)$ belongs to the class ${\cal C}$ if 
\begin{equation}
\int_{-\infty}^{+\infty} \frac{\log^+ |f(t)|}{1+t^2} d t < \infty, \label{eq:classC}
\end{equation}
where the function $\log^+ |w|$ is defined as $\log^+ |w|= \max (0,\log|w|)$ or equivalently as $\log^+ |w|= 1/(2 \pi) \int_0^{2 \pi} \log |w-\exp(\imath \theta)|d \theta$.

A function $f(t)$ bounded on the real axis satisfies $\log^+ |f(t)| \le |f(t)|$. To prove that $f(z)$ belongs to the class ${\cal C}$ it is sufficient to prove that $|f(t)|$ is bounded for every $t \in \Rmath$ and belongs to class ${\cal C}$. The radiation impedance given by equation~(\ref{eq:rad-impedance-polar}) can be majored, for $k \in \Rmath$, by:
\begin{eqnarray}
\left| \beta^{mn} (k) \right| & = & \frac{4 }{\pi} \left| \int_0^{\theta_{\alpha}} \left(\int_0^{\frac{a}{\cos \theta}} H_{mn} (r,\theta) e^{\imath k r} d r\right)  d \theta  + \int_{\theta_{\alpha}}^{\pi/2} \left( \int_0^{\frac{b}{\sin \theta}} H_{mn} (r,\theta) e^{\imath k r} d r\right) d \theta \right| \nonumber \\
	& \le & \frac{4 }{\pi} \left( \int_0^{\theta_{\alpha}} \left(\int_0^{\frac{a}{\cos \theta}} \left| H_{mn} (r,\theta) \right| d r\right)  d \theta  + \int_{\theta_{\alpha}}^{\pi/2} \left( \int_0^{\frac{b}{\sin \theta}} \left| H_{mn} (r,\theta) \right| d r \right) d \theta \right). \label{eq:Bmn(k)bound}
\end{eqnarray}
As $H_{mn} (r,\theta)$ is continuous integrable function of $r$ and $\theta$ the two integrals that appear in the right member of equation~(\ref{eq:Bmn(k)bound}) are finite. Then $\beta^{mn} (k)$ is bounded for every $k$ real and belongs to class ${\cal C}$. 

Showing that the resonance function $f^{mn}(k) = k^2 (1-\epsilon \beta^{mn} (k)) + k_{0mn}^2 $ belongs to class ${\cal C}$ requires a more refined estimation as, because of the $k^2$ term, it is unbounded at infinity. Nevertheless, as $\beta^{mn} (k)$ is bounded for every $k$, the resonance function is also bounded for every finite real value $k$, then for every finite value $A$
\begin{equation}
\int_{-A}^{+A} \frac{\log^+ |f^{mn}(t)|}{1+t^2} d t < \infty. \label{int-A+A}
\end{equation} 
It remains to estimate $\int_{\pm A}^{\pm\infty} \frac{\log^+ |f^{mn}(t)|}{1+t^2} d t$. As seen in the previous paragraph, the radiation impedance for real values of $k \rightarrow \infty$ has the following asymptotic approximation 
\begin{equation}
\beta^{mn}(k)  \approx \frac{1}{(\imath k)^4} \left( \int_0^{\theta_{\alpha}} h_{3mn}^1 (\theta) \emath^{\imath k \frac{a}{\cos \theta}} d \theta + \int_{\theta_{\alpha}}^{\frac{\pi}{2}}  h_{3mn}^2 (\theta) \emath^{\imath k \frac{b}{\sin \theta}} d \theta \right). \label{eq:Bmn(k)asympt}
\end{equation}
The two integrals that appear in this equation :
\begin{eqnarray*}
I_1 (k) & = & \int_0^{\theta_{\alpha}} h_{3mn}^1 (\theta) \emath^{\imath k \frac{a}{\cos \theta}} d \theta, \\
I_2 (k) & = &\int_{\theta_{\alpha}}^{\frac{\pi}{2}}  h_{3mn}^2 (\theta) \emath^{\imath k \frac{b}{\sin \theta}} d \theta,
\end{eqnarray*}
can both be approximated for $k \rightarrow \pm \infty$ by the stationary phase method~\cite{Bender-Orzag}. One obtains for $k \rightarrow \pm \infty$
\begin{eqnarray*}
I_1 (k) & \approx & h_{3mn}^1 (0) \sqrt{\frac{\pi}{2 k a}} \emath^{\imath k a + \imath \frac{\pi}{4}} = (-1)^m \frac{m^2 \pi^2}{2 a^3} \sqrt{\frac{\pi}{2 k a}} \emath^{\imath k a + \imath \frac{\pi}{4b}} , \\
I_2 (k) & \approx & h_{3mn}^2 \left(\frac{\pi}{2}\right) \sqrt{\frac{\pi}{2 k b}} \emath^{\imath k b + \imath \frac{\pi}{4}} = (-1)^n \frac{n^2 \pi^2}{2 b^3}\sqrt{\frac{\pi}{2 k b}} \emath^{\imath k b + \imath \frac{\pi}{4}}.
\end{eqnarray*}
Now the radiation impedance equation has the following approximation for $k \rightarrow \pm \infty$
\begin{equation}
\beta^{mn} (k) \approx \frac{1}{(\imath k)^4} \left((-1)^m \frac{m^2 \pi^2}{2 a^3} \sqrt{\frac{\pi}{2 k a}} \emath^{\imath k a + \imath \frac{\pi}{4}} +  (-1)^n \frac{n^2 \pi^2}{2 b^3} \sqrt{\frac{\pi}{2 k b}} \emath^{\imath k b + \imath \frac{\pi}{4}} \right), \label{eq:Bmn(k)stat-phase}
\end{equation}
For $k \rightarrow \pm \infty$, $\beta^{mn} (k)$ tends to zero and the resonance function can be approximated by $f^{mn}(k) \approx k^2$. For $k \rightarrow \pm \infty$ one has $\log^+ |f^{mn}(k)| = \log|k^2|$. Then for every value of $A$
\begin{equation}
\int_{A}^{+\infty} \frac{\log^+ |f^{mn}(t)|}{1+t^2} d t \approx 2 \int_{A}^{+\infty} \frac{\log t}{1+t^2} d t =2  \int_{-A}^{-\infty} \frac{\log |t|}{1+t^2} d t = 2 \frac{1 + \log A}{A} + {\cal O} (1/A^2) <  \infty. \label{int+-A+-infty}
\end{equation}

Combining results of equations~(\ref{int-A+A}) and ~(\ref{int+-A+-infty}) shows that $\int_{-\infty}^{+\infty} \frac{\log^+ |f^{mn}(t)|}{1+t^2} d t < \infty$. Then both impedance radiation and resonance relation belong to the Cartwright class $\cal{C}$.

For $0<\alpha<\pi$ let us denote by $n_{+}(r,\alpha)$ the number of zeros of the function $f^{mn}(z)$ in the sector $\{z : |z| \le  r, |\arg(z)|\le \alpha \}$ and $n_{-}(r,\alpha)$ the number of zeros in $\{z : |z| \le  r, |\pi-\arg(z)| \le \alpha \}$. By the theorem of Cartwright and Levinson (theorem 1, page 127 of~\cite{Levin-1996}) the density of the set of zeros $n_{\pm}(r,\alpha)/r$ has the limit $\lim_{r\rightarrow \infty} n_{\pm}(r,\alpha)/r = \sigma/\pi$. This result is a very strong one, since not only it ensures that the resonance function possesses an infinite number of zeros but also that ``almost all'' zeros lie in the neighborhood of the real axis. 

It is worth noting that the zeros are counted with their multiplicity order. Then the zeros of the resonance function appears as conjugate pairs and then must be counted twice. Actually, the resonance function $f^{mn}(k) = k^2 (1-\epsilon \beta^{mn} (k)) + k_{0mn}^2$ can be approached for $|k|\rightarrow \infty$ by $f^{mn}(k) = k^2 (1-\epsilon 2 h_{3mm}^1 (\theta_1)/k^4  \emath^{\imath k c_1} + k_{0mn}^2$, then if $\hat{k}$ is a solution of $f^{mn}(\hat{k}) = 0$, it is easy to see that $-\hat{k}^{\star}$, the opposite of its conjugate satisfies $f^{mn}(-\hat{k}^{\star}) = 0$. 

\section{Numerical examples}\label{par:num-examples}

In this section, some numerical examples of the multiple resonance phenomenon are presented. One considers a steel ($E=200 \mbox{ Gpa}$, $\nu=0.3$ and $\rho_s = 7800 \mbox{ Kg/m}^3$) simply supported rectangular plate in contact with air (with density $\rho_f = 1.2 \mbox{ Kg/m}^3$ and sound wave celerity $c=340 \mbox{ m/s}$) or water (with density $\rho_f = 1000 \mbox{ Kg/m}^3$ and sound wave celerity $c=1500 \mbox{ m/s}$). The dimensions of the plate are indicated in the chosen examples.

As it is impossible to obtain a closed formula for $J_{1mn}(k,\theta)$ and $J_{2mn}(k,\theta)$ given by equations~(\ref{eq:J1mn}) and~(\ref{eq:J2mn}), a numerically efficient approximation of these functions has been constructed. As shown below, this approximation preserves the order properties of the involved functions. Since one always has $0< \theta_{\alpha}< \pi/2$, the functions to be integrated are expanded with respect to the angular variable $\theta$ to the order 1. This gives
\begin{eqnarray}
J_{1mn}(k,\theta) \approx j_{1m}(k) & = & a \frac{ \imath a^3 k^3  + 2 m^2 \pi^2 - \imath a k m^2 \pi^2 - (-1)^m 2 \emath^{\imath a k} m^2 \pi^2 }{ 4 (a k - m \pi)^2 (a k + m \pi)^2 }, \label{eq:J1mn-approx} \\
J_{2mn}(k,\theta) \approx j_{2n}(k) & = & b \frac{ \imath b^3 k^3 + 2 n^2 \pi^2 - \imath b k n^2 \pi^2 - (-1)^n 2 \emath^{\imath b k} n^2 \pi^2 }{ 4 (b k - n \pi)^2 (n k + n \pi)^2 }. \label{eq:J2mn-approx} 
\end{eqnarray}
After reporting the two expansions (\ref{eq:J1mn-approx}) and (\ref{eq:J1mn-approx}) in the radiation impedance given in equation~(\ref{eq:rad-impedance-polar}), the following approximation of $\beta^{mn} (k)$, denoted by $\check{\beta}^{mn} (k)$, is obtained
\begin{equation}
\check{\beta}^{mn} (k) = -\frac{4}{\pi} ( j_{1m}(k) \theta_{\alpha} + j_{2n}(k)(\pi/2-\theta_{\alpha})). \label{eq:Bmn(k)approx}
\end{equation}
It is worth noting that the accuracy of the approximation of $\beta^{mn} (k)$ decreases as the mode order increases. However, at low orders (say $m \le 3$, $n \le 3$) it remains sufficient. It can easily be see that
\begin{eqnarray}
j_{1m}(k=\pm m\pi/a) & = & a \frac{ \pm 3 \imath + m \pi }{ 16 m \pi }, \label{eq:j_{1m}(mpia)}\\
         j_{1m}(k=0) & = & a \frac{ 1- (-1)^m }{2 m^2 \pi^2 }, \nonumber \\
j_{2n}(k=\pm n\pi/b) & = & b \frac{ \pm 3 \imath + n \pi }{ 16 n \pi }, \label{eq:j_{2n}(npib)}\\
         j_{2n}(k=0) & = & b \frac{ 1- (-1)^n  }{2 n^2 \pi^2 }. \nonumber 
\end{eqnarray}
Then as $j_{1m}(k)$ and $j_{2n}(k)$ are entire functions of $k$, $\check{\beta}^{mn} (k)$ is also an entire function of $k$. It still remains to estimate the order of $\check{\beta}^{mn} (k)$. For given $m$ and $n$ and for $|k| \rightarrow \infty$, one has
\begin{eqnarray*}
\lim_{|k|\rightarrow \infty} j_{1m}(k) & = & \frac{ - (-1)^m m^2 \pi^2 \emath^{\imath a k}  }{ 2 a^3 k^4 }, \\
\lim_{|k|\rightarrow \infty} j_{2n}(k) & = & \frac{ - (-1)^n n^2 \pi^2 \emath^{\imath b k}  }{ 2 b^3 k^4 },
\end{eqnarray*}
and  
\begin{equation}
\lim_{|k|\rightarrow \infty} \check{\beta}^{mn} (k) = 2 \pi \left( (-1)^m  \frac{m^2}{a^3} \theta_{\alpha} \emath^{\imath a k} + (-1)^n \frac{n^2}{b^3} \left(\frac{\pi}{2}- \theta_{\alpha} \right) \emath^{\imath b k} \right) \frac{ 1  }{ k^4 }. \label{eq:Bmn(k)limit}
\end{equation}
It is worth noting that the approximation of the radiation impedance given in equation (\ref{eq:Bmn(k)approx}) or in equation (\ref{eq:Bmn(k)limit}) presents a similar comportment than the exact one: a sum of exponential functions divided by a polynomial of the fourth degree. To confirm that the approximation keeps the order of $\check{\beta}^{mn} (k)$ to one, classical results on the sum and product of entire functions are used again. First it is worth noting that the entire function $(-1)^m  m^2/a^3 \theta_{\alpha} \emath^{\imath a k} + (-1)^n n^2/b^3 \left(\pi/2- \theta_{\alpha} \right) \emath^{\imath b k}$ cannot be identically zero for any value of $a$, $b$, $m$ or $n$ when $|k| \rightarrow \infty$. To see this more clearly, one can deal separately with the cases $a=b$ and a$\neq b$.

\paragraph{The case $a = b$.}
One has $\theta_{\alpha} = \pi/4$ and $\check{\beta}^{mn} (k) \approx \frac{\pi^2}{2 a^3} \left( (-1)^m m^2 + (-1)^n n^2 \right) \emath^{\imath a k} \frac{ 1 }{ k^4 }$ becomes the product of the exponential function $\emath^{\imath a k}$ of order 1 and the inverse of a degree 4 polynomial in $k$, which is of order 0. $\check{\beta}^{mn} (k)$ is therefore an entire function of order 1 and of type $a$. 

\paragraph{The case $a \neq b$.}
Let us take for example $a>b$. Then with $|k| \rightarrow \infty$ one has $\max \left|\emath^{\imath a k}\right| \gg \max \left|\emath^{\imath b k}\right|$; it is  worth noting that this is true in the half plane $\Im (k) <0$, whereas in the half plane $\Im (k) > 0$, $\check{\beta}^{mn} (k)$ tend rapidly to zero. In the half plane $\Im (k) <0$, one therefore has $\check{\beta}^{mn} (k) \approx \frac{2\pi}{a^3} \left( (-1)^m m^2 \theta_{\alpha} \emath^{\imath a k} \right) \frac{ 1  }{ k^4 }$ which is also the product of the exponential function $\emath^{\imath a k}$ of order 1 and the inverse of a degree 4 polynomial in $k$, which is of order 0. $\check{\beta}^{mn} (k)$ is therefore an entire function of order 1 and of type $a$. It is worth noting that if $b>a$ its  type is $b$. In other words, one can assume that $\sigma = \max(a,b)$.

\begin{figure}
\centering
\includegraphics[width=7cm,height=7cm]{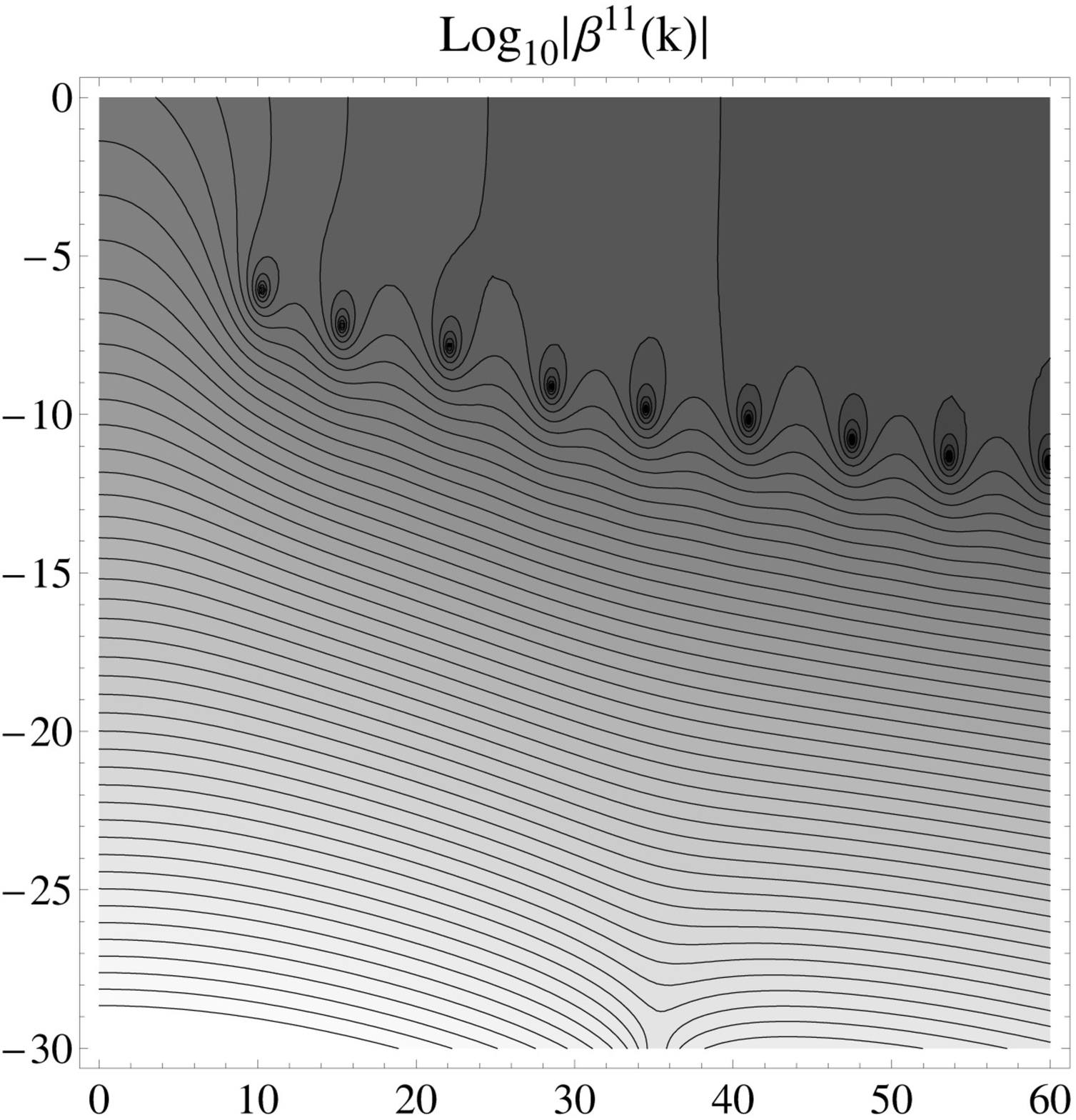} \includegraphics[width=7cm,height=7.2cm]{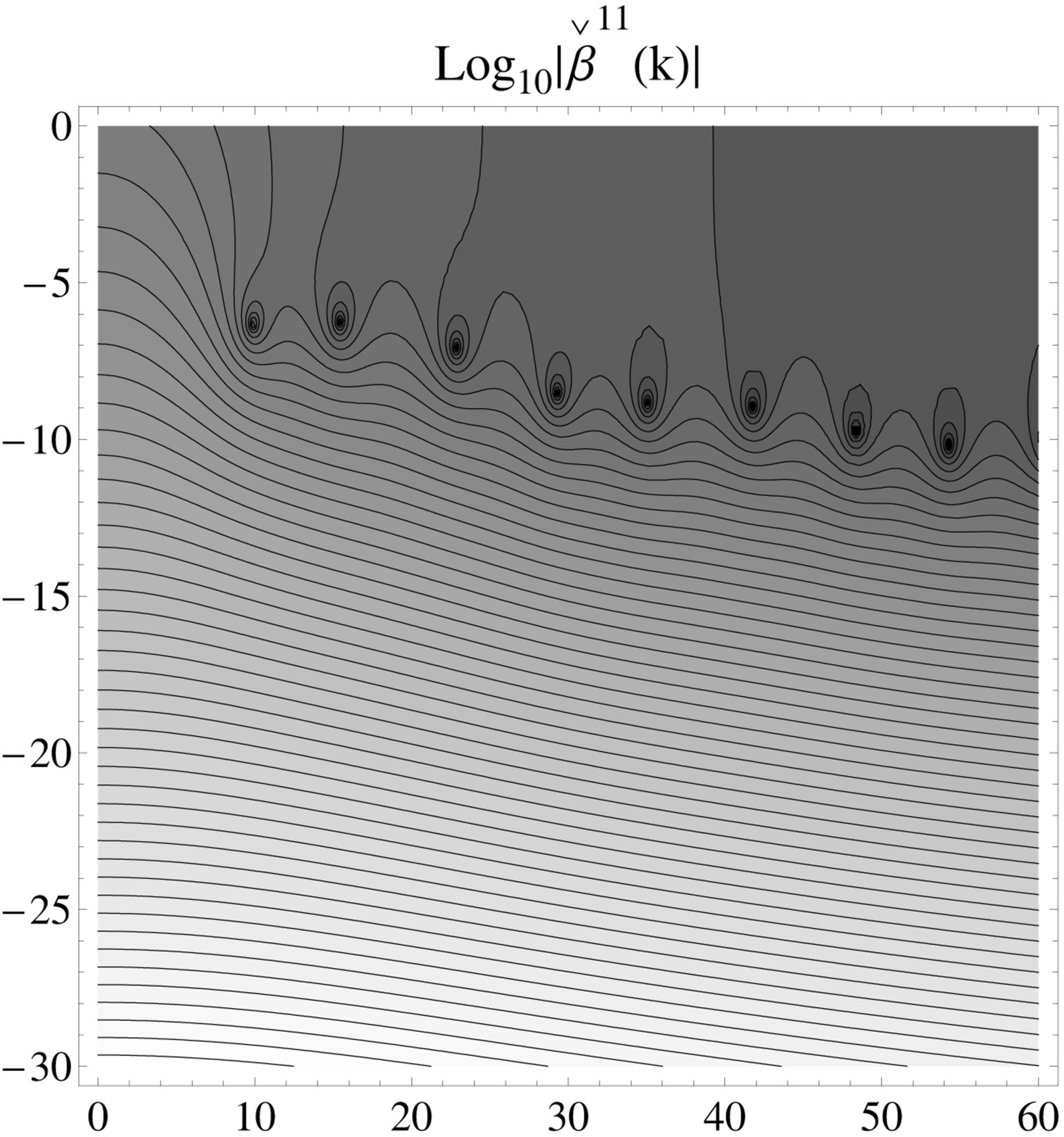}
\caption{Contour plots of the amplitude of the radiation impedance in the mode $m=1$, $n=1$ with $a=1 \mbox{ m }$ and $b=0.7\mbox{ m }$. Left: exact $\beta^{11} ( k)$, right: approximate $\check{\beta}^{11}( k)$. } \label{fig:beta_mode11}
\end{figure}

\begin{figure}
\centering
\includegraphics[width=7cm,height=7cm]{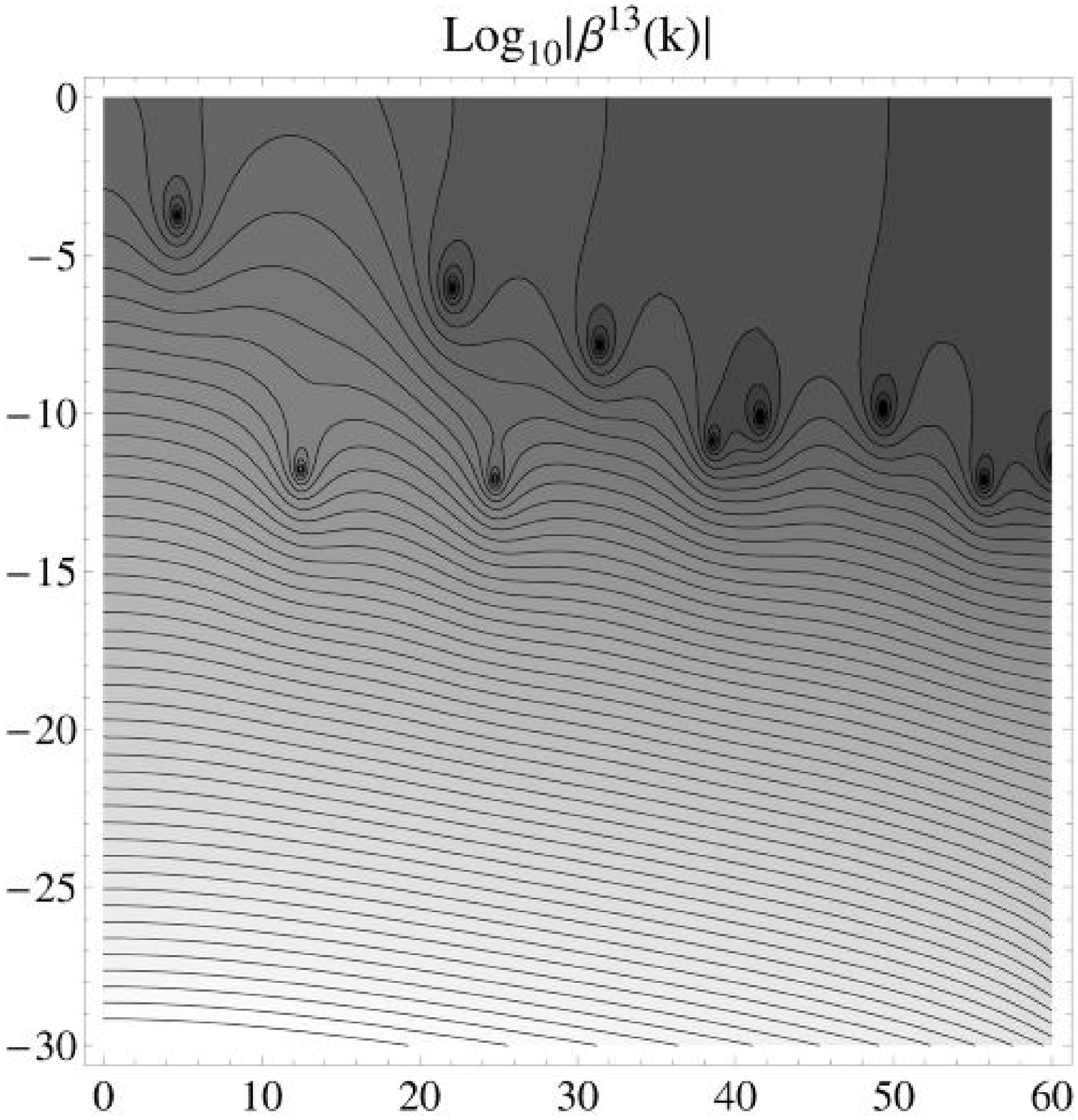} \includegraphics[width=7cm,height=7.2cm]{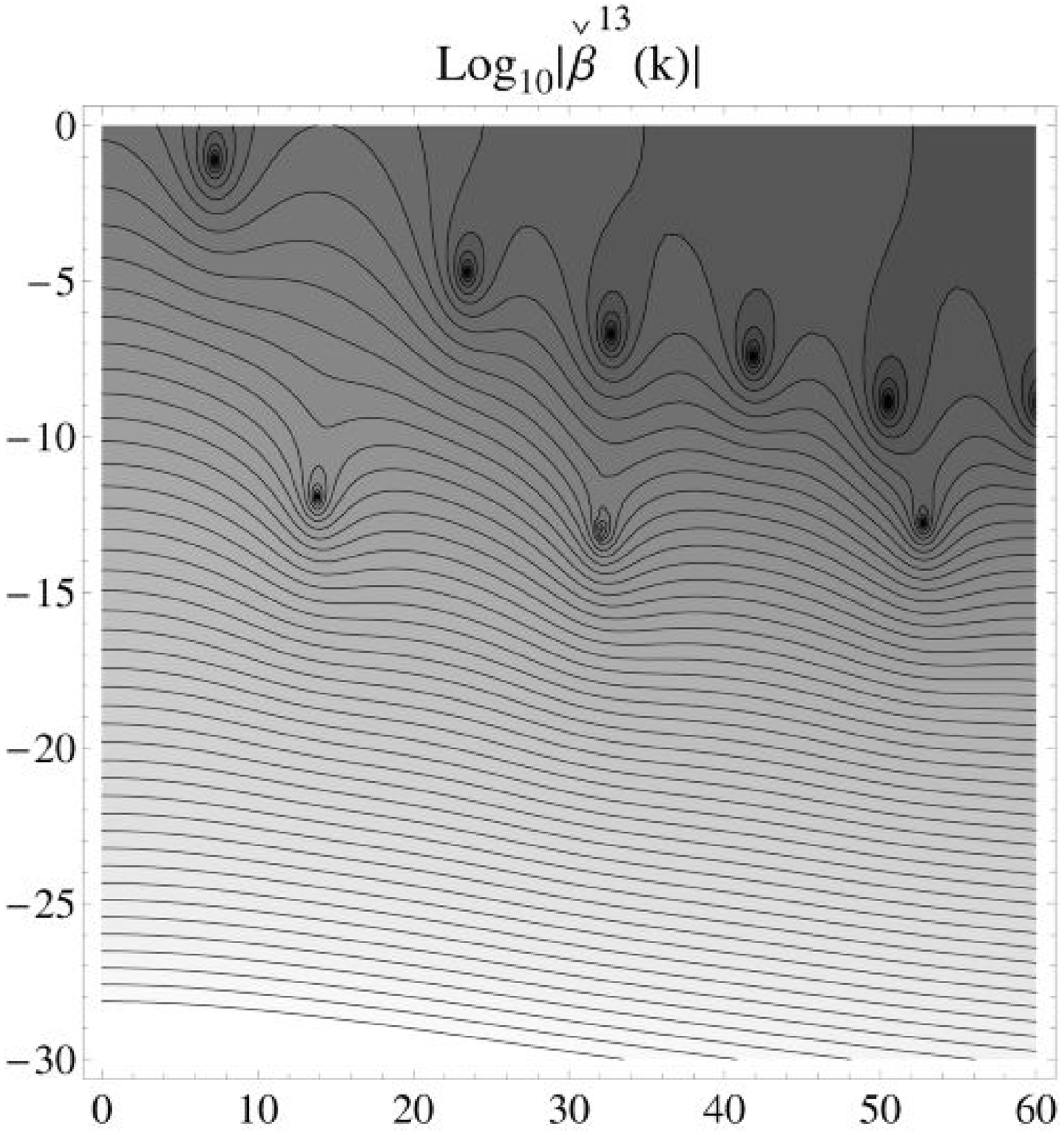}
\caption{Contour plots of the amplitude of the radiation impedance in the mode $m=1$, $n=3$ with $a=1 \mbox{ m }$ and $b=0.7\mbox{ m }$. Left: exact $\beta^{13} ( k)$, right: approximate $\check{\beta}^{13}( k)$. } \label{fig:beta_mode13}
\end{figure}

Figures~(\ref{fig:beta_mode11}) and~(\ref{fig:beta_mode13}) give contour plots for the amplitude of the radiation impedance $\beta^{mn}( k)$ and its app:oximation $\check{\beta}^{mn} (k)$ in the mode $m=1$ $n=1$ in figure~(\ref{fig:beta_mode11}) and in the mode $m=1$ $n=3$ in figure~(\ref{fig:beta_mode13}). In these curves, the dark spots correspond to the zeros of the radiation impedances. It is worth noting that the lower the mode order, the better the approximation $\check{\beta}^{mn} (k)$ becomes.

Both radiation impedance $\check{\beta}^{mn} (k)$ and resonance relation $\check{f}^{mn}(k)$ belong to the Cartwright class $\cal C$ and then all, but a finite number of, their zeros must lie in the neighborhood of the real axis. 

As shown above the type $\sigma$ of $\check{\beta}^{mn} (k)$ is given by $\sigma =\max(a,b)$. Then the type of $\check{f}^{mn}(k)$ is given the type of $k^2 (1-\epsilon \beta^{mn} (k)) + k_{0mn}^2$; and for $\epsilon \neq 0$ it is given by $\sigma =\max(a,b)$. By the Cartwright and Levinson theorem, the density of resonances $\lim_{r\rightarrow \infty} (n_{+}(r,\alpha)/r + n_{-}(r,\alpha)/r )$ is then given by $2\max(a,b)/\pi$.

To verify this, the simplest way is to count the zeros of $f^{mn}(z)$ using the argument theorem on the disk C with radius K centered on the origin. It is clear from equations~(\ref{eq:j_{1m}(mpia)}) and~(\ref{eq:j_{2n}(npib)}) that $\check{\beta}^{mn}( k)$ and  $\check{f}^{mn}(k)$ have no pole inside $C$. The number of its zeros inside this disk, in the case of the mode $(m-n)$, is denoted by $n_{\check{f}^{mn}}(K)$. It is given by the integral $n_{\check{f}^{mn}}(K) = 1/(2 \imath \pi) \oint_{C} \check{f}^{\prime mn} (k)/ \check{f}^{mn} (k) d k$ with $k = K e^{\imath \theta}$. This integral is computed numerically without difficulty. As shown before, the zeros occur in combined pairs of negative imaginary part, then each zero must be counted twice; then the density of the zeros given by $(n(K,\alpha)/K =  n_{+}(K,\alpha)/K + n_{-}(K,\alpha)/K$  has the asymptotic value $\lim_{K\rightarrow \infty} n(K,\alpha)/ K = 2 n_{\check{f}^{mn}}(K)/K$. 

For example, let us take air for the fluid and let the plate dimensions be $a = 1 \mbox{ m}$, $b = 0.7 \mbox{ m}$ and thickness $h = 0.002 \mbox{ m}$. It can easily be established that in the mode $m = 1$, $n = 1$, one has $n_{\check{f}_{11}} (0.3) = 2$, $n_{\check{f}_{11}} (12) = 2$ and $n_{\check{f}_{11}} (30) = 8$. In the mode $m = 1$, $n = 3$, one has $n_{\check{f}_{13}} (2) = 2$, $n_{\check{f}_{13}} (10) = 2$, $n_{\check{f}_{13}} (30) = 8$. As the radius $K$ of the disk $C$ becomes large one obtains for the density $n_{\check{f}^{mn}}(K)/K$ the following values

\begin{eqnarray*}
& & a=1\mbox{ m}, b=0.7\mbox{ m}, K = 20; \frac{n_{\check{f}^{11}}(K)}{K} = 1/3.3333 \\
& & a=1\mbox{ m}, b=0.7\mbox{ m}, K = 200; \frac{n_{\check{f}^{11}}(K)}{K} = 1/3.125 \\
& & a=1\mbox{ m}, b=0.7\mbox{ m}, K = 20 000; \frac{n_{\check{f}^{11}}(K)}{K} = 1/3.14159 \\
& & a=0.5\mbox{ m}, b=0.5\mbox{ m}, K = 200 00; \frac{n_{\check{f}^{11}}(K)}{K} = 1/6.28319 \\
& & a=0.5\mbox{ m}, b=0.1\mbox{ m}, K = 200 00; \frac{n_{\check{f}^{11}}(K)}{K} = 1/6.28319  \\
& & a=0.1\mbox{ m}, b=0.1\mbox{ m}, K = 200 00; \frac{n_{\check{f}^{11}}(K)}{K} = 1/31.4159 \\
& & a=1\mbox{ m}, b=0.7\mbox{ m}, K = 200 00; \frac{n_{\check{f}^{12}}(K)}{K} = \frac{n_{\check{f}^{15}}(K)}{K} = \frac{n_{\check{f}^{32}}(K)}{K} = 1/3.14159 
\end{eqnarray*}

This shows that $\lim_{K\rightarrow \infty} n(K,\alpha)/ K  = 2 n_{\check{f}^{mn}}(K) / K  = 2 \max(a,b)/\pi$. All these numeric values are a validation of the theoretical results obtained in the paragraph~\ref{par:multiple}. 

\begin{figure}
\centering
\includegraphics[width=7cm,height=7cm]{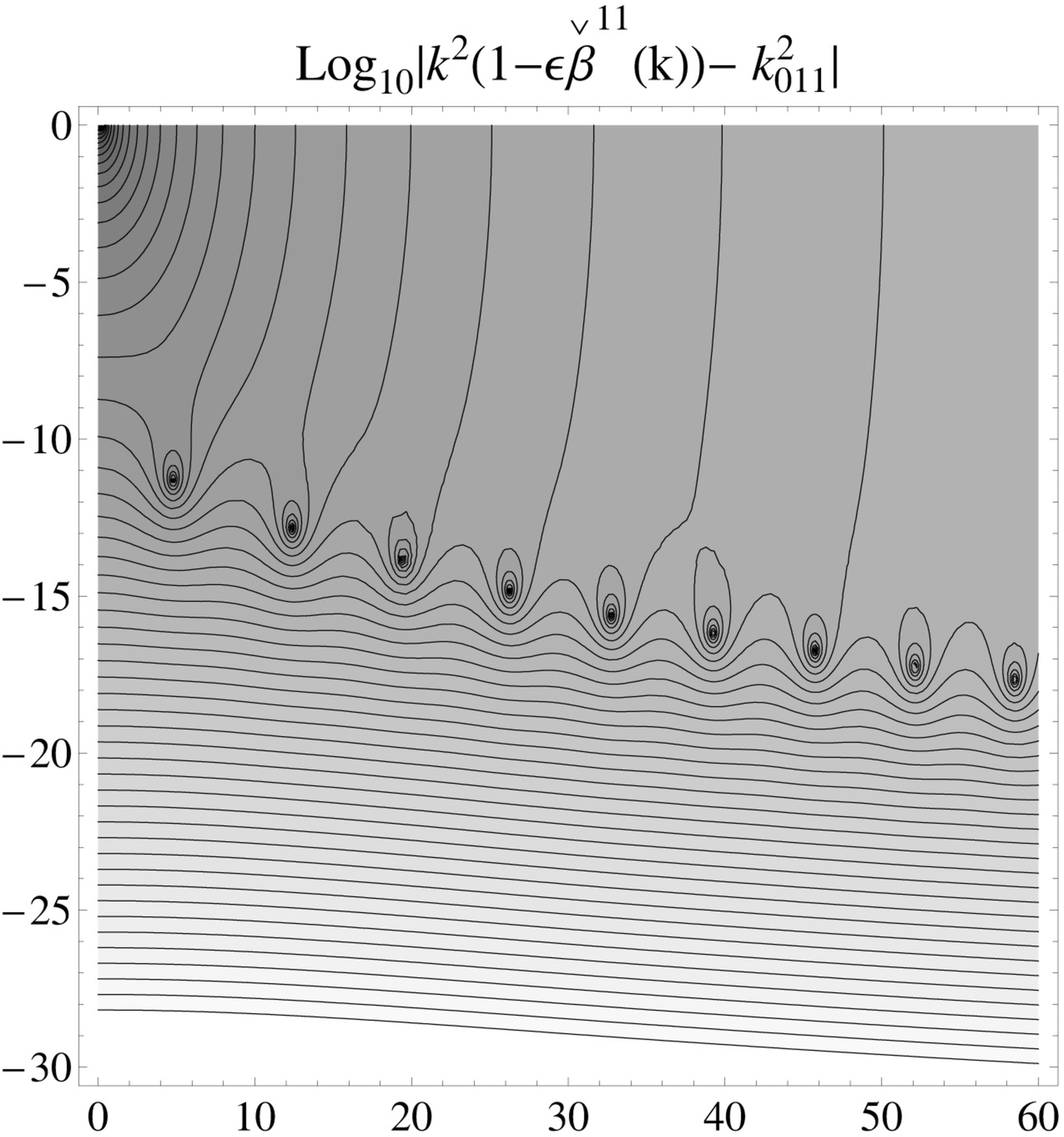} \includegraphics[width=7cm,height=7cm]{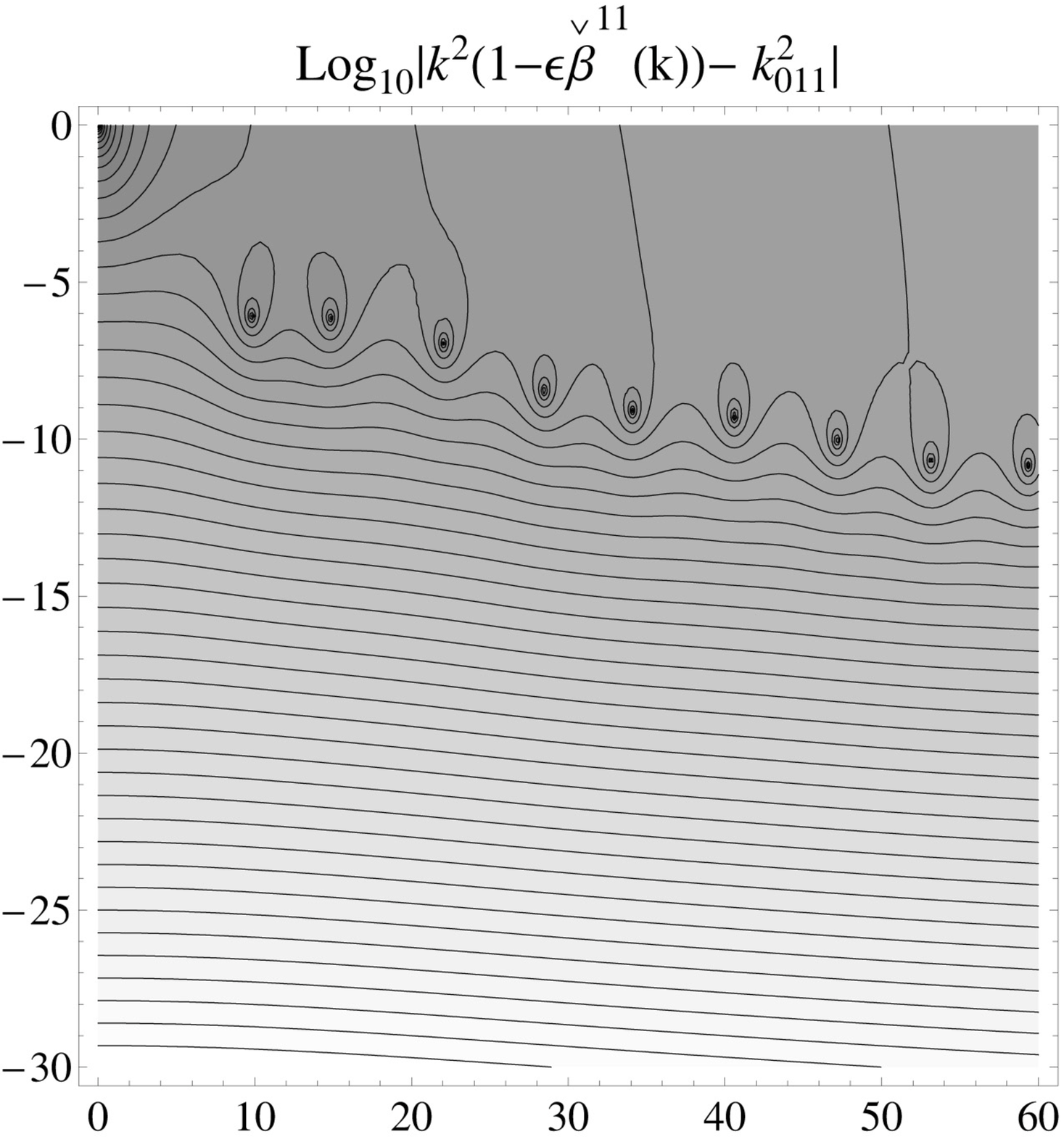}
\caption{Contour plots of the amplitude of the resonance value equation $k^2 (1 - \epsilon \check{\beta}^{mn}(k)) - k_{0mn}^{2}$ in the mode $m=1$, $n=1$ in the case with $a=1 \mbox{ m }$, $b=0.7\mbox{ m }$ and $h = 2 \mbox{ mm}$. Left: steel plate in contact with air ($\epsilon \approx 0.08$), right: steel plate in contact with water ($\epsilon \approx 64$). } \label{fig:eigenvalue_2mm_11_air_water}
\end{figure}

\begin{figure}
\centering
\includegraphics[width=7cm,height=7cm]{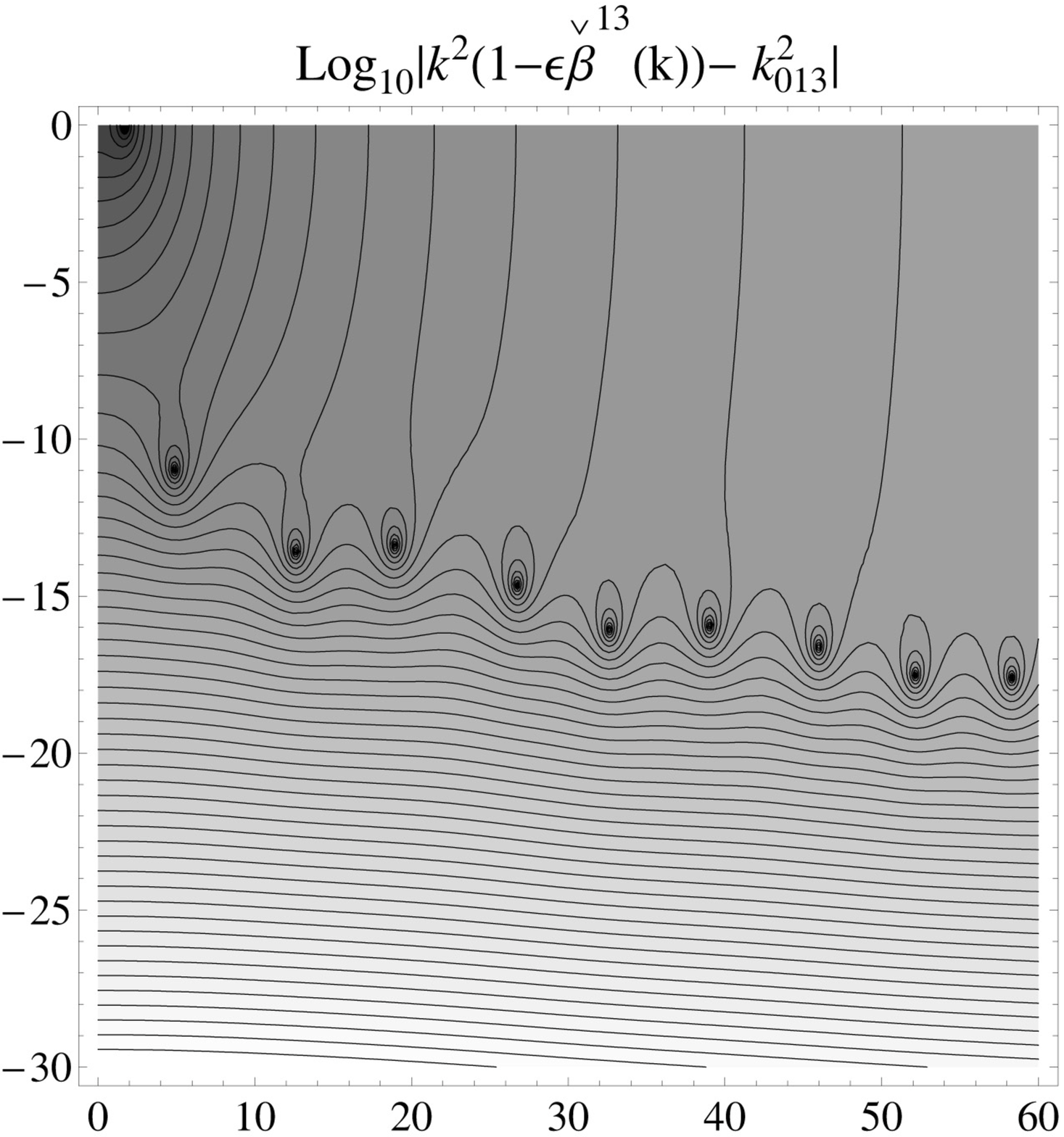} \includegraphics[width=7cm,height=7cm]{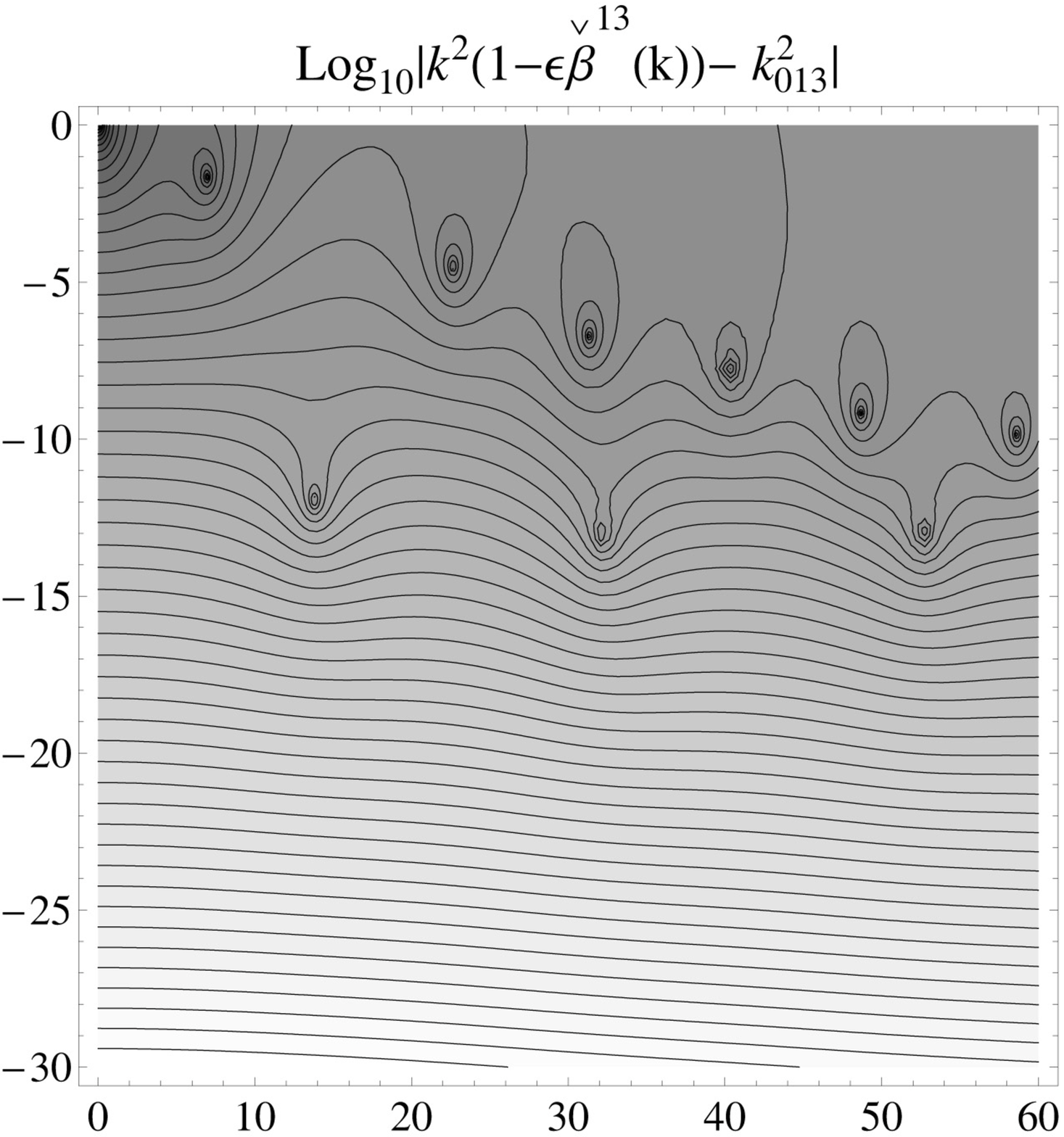}
\caption{Contour plots of the amplitude of the resonance value equation $k^2 (1 - \epsilon \check{\beta}^{mn}(k)) - k_{0mn}^{2}$ in the mode $m=1$, $n=3$ in the case with $a=1 \mbox{ m }$, $b=0.7\mbox{ m }$ and $h = 2 \mbox{ mm}$. Left: steel plate in contact with air ($\epsilon \approx 0.08$), right: steel plate in contact with water ($\epsilon \approx 64$). } \label{fig:eigenvalue_2mm_13_air_water}
\end{figure}

The resonance relation of the fluid loaded plate has an infinite number of zeros. Therefore, while in a vacuum, each mode of a simply supported plate will have only one resonance frequency (and its opposite), under loading condition each mode will have an infinite number of resonance frequencies. This result is confirmed in figures~\ref{fig:eigenvalue_2mm_11_air_water} and~\ref{fig:eigenvalue_2mm_13_air_water} where the dark spots that appear in the curves correspond to the zeros of the approximate resonance value relation $\check{f}_{mn} (k)=0$. The results presented in figures~\ref{fig:eigenvalue_2mm_11_air_water} and~\ref{fig:eigenvalue_2mm_13_air_water} also show that except for the zero close to the real axis, the zeros of the resonance value relation corresponding to the plate loaded with water are very similar to those given by the radiation impedance.

Now let us return to the initial question of the existence or non-existence of multiple resonances in fluid-loaded plates. In every  boundary condition, one has to look for the resonance pulsations $\hat{\omega}_{mn}$ such that $\hat{\omega}_{mn}^2 (1 - \epsilon \beta_{\hat{\omega}_{mn} }^{mn}(\omega)) - \tilde{\omega}_{mn}^{(0)2} = 0$. Although it has not been proved that such a comportment is true under various boundary conditions, it seems reasonable to assume that under boundary conditions of all kinds every resonance mode in a plate has an infinite number of resonance frequencies as $\epsilon \neq 0$ (the radiation impedance given by relation~(\ref{eq:rad-impedance}) seems to be an EFET for a large variety of functions $G_{m}$ and $G_{n}$). 

\begin{figure}
\centering
\includegraphics[width=7cm,height=7cm]{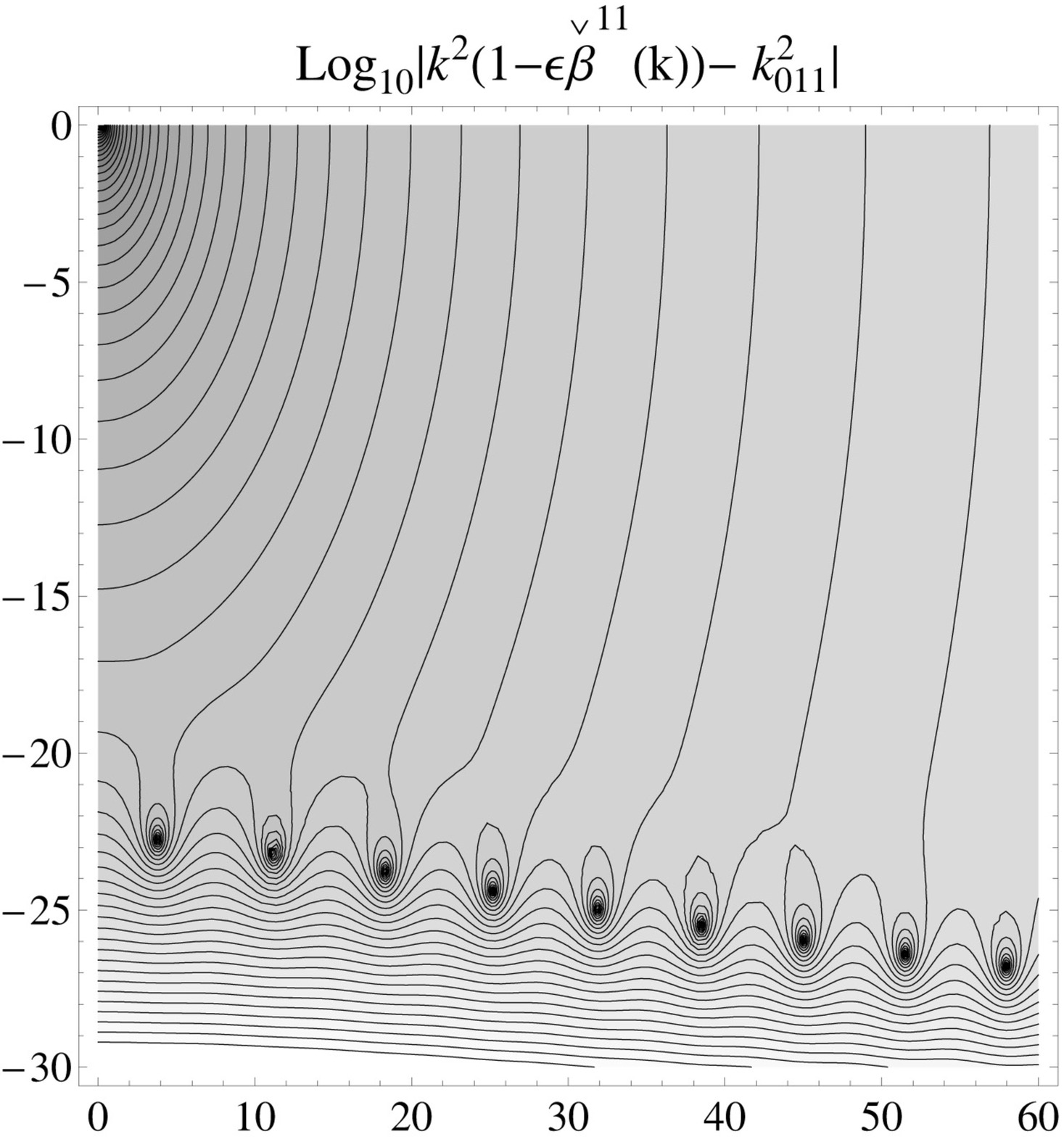} \includegraphics[width=7cm,height=7cm]{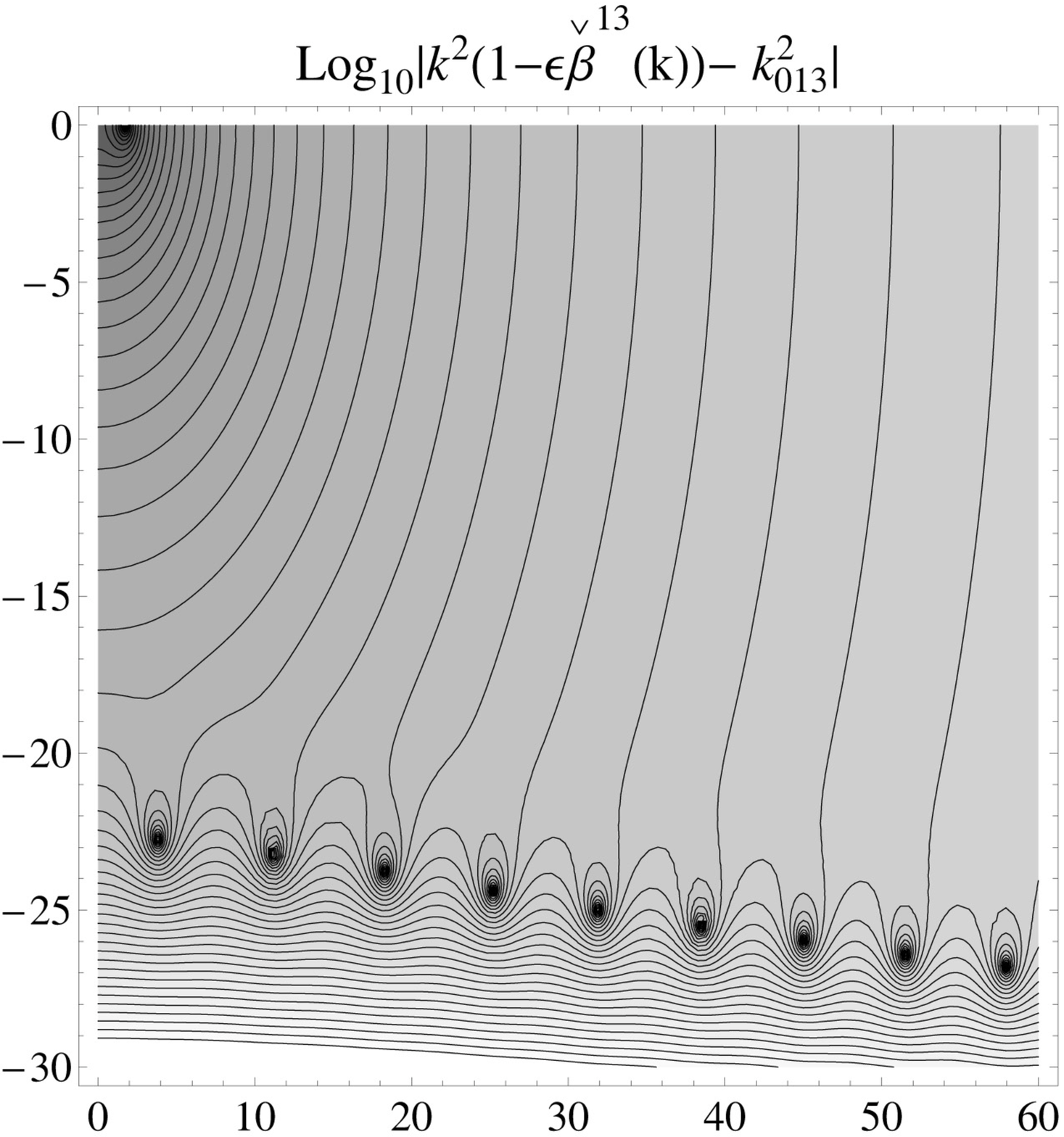}
\caption{Contour plots of the amplitude of the resonance value equation $k^2 (1 - \epsilon \check{\beta}^{mn}(k)) - k_{0mn}^{2}$ in the case $\epsilon=10^{-5}$ with $a=1 \mbox{ m }$, $b=0.7\mbox{ m }$ and $h = 2 \mbox{ mm}$. Left: in the mode $m=1$, $n=1$, right: in the mode $m=1$, $n=3$.} \label{fig:eigenvalue_2mm_11_13_eps10-5}
\end{figure}

As shown by Sanchez~\cite{Sanchez-Sanchez-1989}, under very light loading condition (as in the case of a thick steel plate in contact with air, for example), there exist only the resonances of the structure alone (one in each mode shifted toward the complex plane) and the complex scattering resonances of the fluid in contact with the rigid solid. This contradictory situation is only an apparent one since for each resonance one has a root located very close to imaginary axis (that corresponds to the resonance of the in vacuo mode shifted through the complex plane by the coupling with the fluid with added mass and added damping) and an infinite series of roots corresponding to the zeros of the radiation impedance. For $\epsilon \rightarrow 0$, that is the case considered by Sanchez, all resonances sets introduced by the radiation impedance of each mode merge together and form a unique set of resonances for the resonance function. The demonstration seems very difficult but a numerical example of this merging is showed in the figure~\ref{fig:eigenvalue_2mm_11_13_eps10-5} in the modes $m=1$, $n=1$ and $m=1$ $n=3$ in the arbitrary case $\epsilon= 10^{-5}$. In this case, the structure has only two sets of resonances in accordance to the results by Sanchez, the first one is the set of resonances of the structure that are very close to the real axis plane (composed by the set of resonances of each modes of the structure with small imaginary part) and the second which is the resonance set of the rigid structure alone (corresponding to the merged zeros of the radiation impedance with large negative imaginary part).

Now under very heavy loading condition (as in the case of a very thin plate in contact with heavy density fluid ), what is the relevant relation ? This seems to be  an open question. Although the resonance relation $k^2 (1 - \epsilon \beta^{mn}(k)) -  k_{mn}^{(0)2} = 0$ gives good approximation to the spectrum under heavy loading~\cite{POM-2007}, it has to be supplemented by terms accounting for the cross modal impedance, which make it almost impossible to conduct the above analysis.

%==================================================================
 
\section{Conclusions} \label{par:conclusions}
The aim of this paper was to show that the loading of a vibrating plate with a non-vanishing density fluid transforms each resonance of the structure into an infinite number of resonances. This result opens up new prospects. In particular, it would be interesting to obtain similar estimates on the vibratory modes occurring in plates with various boundary conditions, shape or loading (e.g. for non baffled plate) or in mechanical structures of other kinds, although these may not be easy to obtain. Another question which arises is how to determine under what loading conditions (moderate or heavy) a coupled system, such as a thin steel plate in contact with water, will have these properties. Another important aspect worth investigating is the validity of modal expansions in the case of resonance modes having two or more resonances frequencies such as those studied in this paper. It would also be interesting to develop experimental device which would make it possible to test the occurrence of the behavior described here.

%==================================================================

\appendix

\section{Derivatives of the function $H_{mn}(r,\theta)$ with respect to $r$}

The function $H_{mn}(r,\theta)$ is given by

\begin{eqnarray*}
H_{mn}(r,\theta) = \frac{1}{4 m n \pi ^2} &\!\!\! & \left[m \pi \left(1-\frac{r \cos \theta }{a}\right) \cos \left(\frac{m \pi r \cos \theta }{a}\right) + \sin \left(\frac{m \pi  r \cos \theta }{a}\right)\right] \\
&\!\!\! & \left[n \pi \left(1-\frac{r \sin \theta}{b}\right) \cos\left(\frac{n \pi r \sin \theta}{b}\right) +\sin \left(\frac{n \pi r \sin theta}{b}\right)\right].
\end{eqnarray*}

A tedious calculation (done without difficulty under any symbolic program) shows that the derivatives with respect to $r$ of $H_{mn}(r,\theta)$ up to order 3 taken at the points $r=0$, $r=a/\cos \theta$ and $r=b/\sin \theta$ are given by :
\begin{eqnarray*}
\left[ H_{mn}(r,\theta) \right]_{r=0} & = & \frac{1}{4}  \\
\left[ H_{mn}(r,\theta) \right]_{r=\frac{a}{\cos \theta} } & = & 0  \\
\left[ H_{mn}(r,\theta) \right]_{r=\frac{b}{\sin \theta} } & = & 0  \\
\left[ \frac{\partial H_{mn}(r,\theta)}{\partial r} \right]_{r=0} & = & 0  \nonumber \\
\left[ \frac{\partial H_{mn}(r,\theta)}{\partial r} \right]_{r=\frac{a}{\cos \theta} } & = & 0 \\
\left[ \frac{\partial H_{mn}(r,\theta)}{\partial r} \right]_{r=\frac{b}{\sin \theta} } & = & 0  \\
\left[ \frac{\partial^2 H_{mn}(r,\theta)}{\partial r^2} \right]_{r=0} & = & -\frac{\pi ^2 \left(b^2 m^2 \cos^2 \theta +a^2 n^2 \sin^2 \theta \right)}{4 a^2 b^2} \\
& = & h_{2mn}(\theta) \nonumber \\
\left[ \frac{\partial^2 H_{mn}(r,\theta)}{\partial r^2} \right]_{r=\frac{a}{\cos \theta} } & = & 0  \\
\left[ \frac{\partial^2 H_{mn}(r,\theta)}{\partial r^2} \right]_{r=\frac{b}{\sin \theta} } & = & 0   \\
\left[ \frac{\partial^3 H_{mn}(r,\theta)}{\partial r^3} \right]_{r=0} & = & \frac{1}{2} \pi ^2 \left(\frac{m^2 \cos ^3 \theta}{a^3}+\frac{n^2 \sin^3 \theta}{b^3} \right)  \\
& = & h_{3mn}^0 (\theta) \\
\left[ \frac{\partial^3 H_{mn}(r,\theta)}{\partial r^3} \right]_{r=\frac{a}{\cos \theta} } & = & \frac{(-1)^m m^2 \pi \cos^2 \theta}{2 a^3 b n} \left[ b \cos \theta \left(n \pi  \cos \left(\frac{a n \pi \tan \theta}{b}\right)+\sin \left(\frac{a n \pi \tan \theta}{b}\right)\right) \right. \\ 
& & \hspace{8em} \left. -a n \pi \cos \left(\frac{a n \pi \tan\theta}{b}\right) \sin \theta \right]\\
& = & h_{3mn}^1 (\theta) \\
\left[ \frac{\partial^3 H_{mn}(r,\theta)}{\partial r^3} \right]_{r=\frac{b}{\sin \theta} } & = & \frac{(-1)^n n^2 \pi  sin^2\theta}{2 a b^3 m} \left[a \sin \theta \left(m \pi \cos \left(\frac{b m \pi \cot \theta}{a}\right)+\sin \left(\frac{b m \pi \cot \theta}{a}\right)\right) \right. \\ 
& & \hspace{8em} \left. -b m \pi  \cos \theta \cos \left(\frac{b m \pi \cot \theta}{a}\right)\right]  \\
& = & h_{3mn}^2 (\theta)
\end{eqnarray*}

\end{document}